\documentclass{llncs}
\usepackage{graphicx}

\begin{document}

\title{Practical Random Access\\to SLP-Compressed Texts}
\author{Travis Gagie\inst{1}\orcidID{0000-0003-3689-327X} \and Tomohiro I\inst{2}\orcidID{0000-0001-9106-6192} \and Giovanni Manzini\inst{3}\orcidID{0000-0002-5047-0196} \and Gonzalo Navarro\inst{4}\orcidID{0000-0002-2286-741X} \and Hiroshi Sakamoto\inst{2}\orcidID{0000-0002-3470-9187} \and Louisa Seelbach Benkner\inst{5}\orcidID{0000-0002-3204-3801} \and Yoshimasa Takabatake\inst{2}\orcidID{0000-0002-4566-8974}}

\institute{Dalhousie University, Halifax, Canada \and
Kyushu Institute of Technology, Fukuoka, Japan \and
University of Eastern Piedmont, Alessandria, Italy \and
CeBiB \& DCC, University of Chile, Santiago, Chile \and
University of Siegen, Siegen, Germany}

\maketitle

\begin{abstract}
Grammar-based compression is a popular and powerful approach to compressing repetitive texts but until recently its relatively poor time-space trade-offs during real-life construction made it impractical for truly massive datasets such as genomic databases.  In a recent paper (SPIRE 2019) we showed how simple pre-processing can dramatically improve those trade-offs, and in this paper we turn our attention to one of the features that make grammar-based compression so attractive: the possibility of supporting fast random access.  This is an essential primitive in many algorithms that process grammar-compressed texts without decompressing them and so many theoretical bounds have been published about it, but experimentation has lagged behind.  We give a new encoding of grammars that is about as small as the practical state of the art (Maruyama et al., SPIRE 2013) but with significantly faster queries.
\end{abstract}

\section{Background}
\label{sec:background}

It is widely acknowledged that we now have more data than we can properly handle, and one possible solution is to compress it in such a way that we can later process it quickly without decompressing it.  Since many of our largest and most important datasets --- such as genomic databases --- are highly repetitive texts, grammar-based schemes offer excellent compression ratios while still admitting algorithms for many natural problems that run in times polynomial in the sizes of the compressed representations.

Probably the most popular such schemes are those producing straight-line programs (SLPs), which are context-free grammars in Chomsky normal form that each generate exactly one string; we refer the reader to Lohrey's~\cite{Lohrey12} and Navarro's~\cite{Nav20} surveys for more details of SLPs, SLP algorithmics, SLP-based data structures, and related techniques.  Since many algorithms that process SLPs depend on random access to the compressed texts as a primitive operation, there have been several important theoretical papers written about supporting it, which we review in Appendix~\ref{app:theory}.

Unfortunately, there have not been as many breakthroughs about supporting random access to SLP-compressed texts in practice.  Block trees~\cite{belazzougui2015queries} are practical, and resemble SLPs in many ways with similar theoretical bounds, but they are not SLPs nor even context-free grammars and so researchers studying SLP algorithmics may wish to avoid them.  Variant call format~\cite{danecek11} and relative Lempel-Ziv~\cite{kuruppu2010relative} are also practical but even less like SLPs.

In the real world, users still rely on Larsson and Moffat's~\cite{LarssonM99} RePair algorithm, even though the SLPs it produces are not optimal in the worst case and it is not known if they are even always close to optimal.\footnote{RePair is probably most commonly used in natural-language processing, where it is viewed as an implementation of Gage's~\cite{Gage94} byte-pair encoding and used for word segmentation in neural machine translation~\cite{SennrichHB16}; we refer readers to Gall\'e's~\cite{Galle19} recent survey for more discussion.}  Similarly, users who need random access to SLP-compressed strings often just augment the SLPs produced by RePair and na\"ively encode them even though, as far as we are aware, there are no good bounds on their heights and thus no good bounds on the access times (unless we modify the SLPs at the risk of making them impractical).  The best encoding we know of is due to Maruyama et al.~\cite{MaruyamaTSS13}, which is significantly smaller than the na\"ive encoding but also significantly slower.

Practitioners' main concern about RePair seems to be the large constants in its time-space trade-offs for construction.  For example, Navarro's implementation of RePair\footnote{\url{https://users.dcc.uchile.cl/~gnavarro/software/repair.tgz}} compresses a 3.0 GB file containing copies of human chromosome 19 from 50 distinct genomes into 23 MB and a 5.9 GB file containing copies from 100 genomes into 24 MB, but on a commodity computer it takes 84 minutes and 11 GB of workspace for the former and 11 hours and 18 GB of workspace for the latter~\cite{GIMNST19}.  Although several alternatives have been proposed~\cite{BGP17,FTNIBK19,GJ17,ohno2018lz,sakai2019repair}, until recently the most practical option for files of more than a few gigabytes was SOLCA~\cite{TIS17}, which compresses the 3.0 GB file into 40 MB using 11 minutes and 310 MB of workspace, and the 5.9 GB file into 45 MB using 22 minutes and 310 MB of workspace, respectively.  In addition to achieving noticeably worse compression than RePair, even SOLCA took over 3.6 hours to compress a 59 GB file containing copies of chromosome 19 from 1000 genomes, although it used only 783 MB of workspace and produced an SLP of only 129 MB.

In a recent paper~\cite{GIMNST19} we showed how simple pre-processing with context-triggered piecewise hashing (CTPH) can dramatically improve the trade-offs for both RePair and SOLCA.  For CTPH, we run a relatively short sliding window over the text and insert a phrase break whenever the Karp-Rabin hash of the window's contents is 0 modulo some parameter $p$.\footnote{We realized after~\cite{GIMNST19} went to press that the worst-case approximation ratios in Theorems 1 and 2 should be multiplied by the length of the sliding window, but this does not affect our approach's correctness or practicality.}  Although it works poorly in the worst case even on repetitive texts --- for example, the string $a^n$ is either parsed into a single phrase or into nearly $n$ of them --- in practice on most repetitive texts CTPH produces a dictionary of distinct phrases and a parse that are, together, much smaller than the text.  We note in passing the similarity of the high-level ideas behind prefix-free parsing and string synchronizing sets~\cite{kempa2019string}, which have good worst-case bounds and seem practical for small files~\cite{Dinklage20} but may not scale as easily to tens or hundreds of gigabytes.

We first experimented with CTPH for building Burrows-Wheeler Transforms (BWT) for massive texts~\cite{boucher2019prefix,kuhnle2019efficient}, because we can quickly build the run-length compressed BWT from the dictionary and the parse in workspace bounded in terms of their combined size.  It then occurred to us that, if we build SLPs for the dictionary and the parse, with the SLP for the dictionary restricted such that each phrase is the complete expansion of some non-terminal, then we can easily combine those SLPs to obtain an SLP for the text: we replace each terminal in the SLP for the parse --- which is a phrase identifier --- by the non-terminal in the SLP for the dictionary whose expansion is that phrase.  For example, on the same commodity computer, applying RePair to the dictionary and parse of the 59 GB file containing 1000 copies of chromosome 19, compressed it by a factor of 1000 in 21 minutes using 7.0 GB of workspace, and applying SOLCA compressed it by a factor of over 400 in 44 minutes using only 4.6 MB of workspace.

Now that grammar-based compression itself is reasonably scalable, it is time to turn our attention to making SLP algorithmics practical, and an obvious starting place is improving the practicality of random access.

\section{Design of the new grammar encoding}
\label{sec:design}

Random access to an SLP-compressed text works by descending the parse tree and computing the expansion sizes of the non-terminals we visit.  In particular, at each non-terminal, we compute the expansion sizes of its children, in order to know to which we should descend.  The main idea of our new encoding is that symbols' expansion sizes can tell us a lot about their identities, so we should tightly integrate how we encode these two kinds of information.

If the non-terminals (excluding the start symbol, unless it expands to two symbols in one step) in an SLP have $d$ distinct expansion sizes, then we build a minimal perfect hash function (MPHF) $h$ that maps those sizes bijectively to the numbers in $[0, d - 1]$.  In this paper we use Esposito, Graf and Vigna's recent RecSplit~\cite{EGV20} MPHF implementation, which occupies only about $1.56 d$ bits.  We note that we cannot recover the $d$ sizes from the MPHF --- given any other size, it will still return a hash value in the range $[0, d - 1]$ --- so in our algorithm we will be careful to query the MPHF only with numbers we know are non-terminals' expansion sizes in our~SLP.

We group the non-terminals by their expansion sizes; sort the groups by the hash values of the expansion sizes of the non-terminals in them; and replace each non-terminal by a triple consisting of the expansion size of its left child, and the offsets of its children in their groups (or, if they are terminals, their offsets in the alphabet).  If the start symbol expands to more than one symbol in one step, then we store a bitvector indicating the lengths of the expansions of the symbols it expands to in one step, and we store the offset of each of those symbols in its group (or its offset in the alphabet if it is a terminal).

The random access to the input text~$T$ works as follows. Suppose we know $T [i]$ is the $j$th character in the expansion of the $k$th non-terminal, say $X$, in the group of non-terminals with expansion size $\ell$.  Using some small auxiliary data structures, we can
\begin{enumerate}
    \item look up $X$'s left child's expansion size $\ell'$;
    \item compute $X$'s right child's expansion size $\ell'' = \ell - \ell'$;
    \item look up $X$'s left child's offset $k'$ in the group of non-terminals with expansion size $\ell'$ (or its offset in the alphabet if $\ell' = 1$ so it is a terminal);
    \item look up $X$'s right child's offset $k''$ in the group of non-terminals with expansion size $\ell''$ (or its offset in the alphabet if $\ell'' = 1$ so it is a terminal);
    \item if $j \leq \ell'$ then set $j' = j$ and recursively find the $j'$th character in the expansion of the $k'$th non-terminal in the group of non-terminals with expansion size $\ell'$ (or just return the character if it is a terminal);
    \item otherwise, $j > \ell'$ and we set $j'' = j - \ell'$ and recursively find the $j''$th character in the expansion of the $k''$th non-terminal in the group of non-terminals with expansion size $\ell''$ (or just return the character if it is a terminal).
\end{enumerate}
Since $T [i]$ is the $(i + 1)$st character in the expansion of the only non-terminal with expansion size $n$, we can descend down the parse tree in time proportional to its height.  If we push the offsets and expansion sizes on a stack as we do so, then we can traverse the parse tree starting from the $(i + 1)$st leaf and thus extract subsequent characters of $T$ in constant amortized time per character.

\begin{figure}[t]
\centering
\begin{tabular}{c@{\hspace{5ex}}c}
\rotatebox{90}
{\includegraphics[height=.7\textwidth]{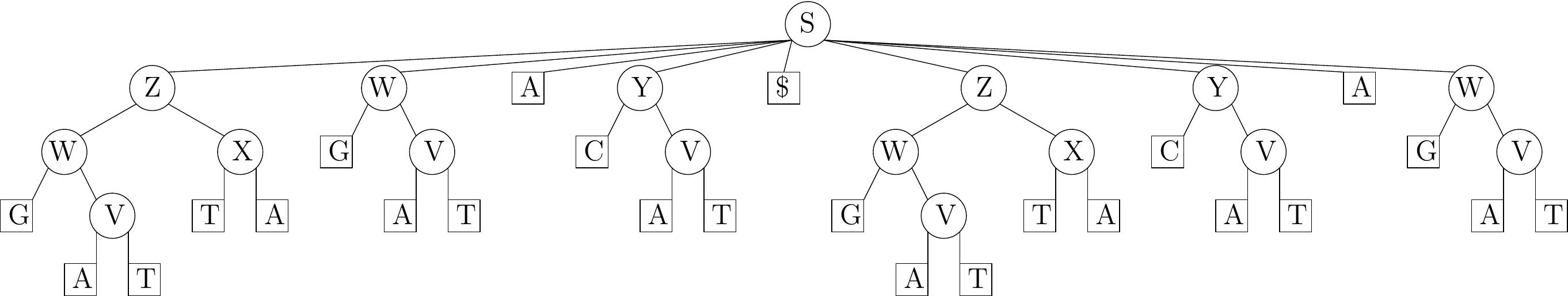}} &
\raisebox{6ex}
{\tiny \begin{tabular}{r@{\hspace{1ex}}l}
S $\rightarrow$ & ZWAY\$ZYAW \\
Z $\rightarrow$ & WX \\
Y $\rightarrow$ & CV \\
X $\rightarrow$ & TA \\
W $\rightarrow$ & GV \\
V $\rightarrow$ & AT
\end{tabular}}
\end{tabular}
\caption{An SLP (right) for GATTAGATACAT\$GATTACATAGAT and its parse tree (left).}
\label{fig:tree}
\end{figure}

\medbreak\noindent{\bf Encoding example.}
Consider the SLP for GATTAGATACAT\$GATTACATAGAT that is shown with its parse tree in Figure~\ref{fig:tree}.  The 3 distinct sizes of the non-terminals' expansions (excluding S) are 5 (for Z), 3 (for W and Y) and 2 (for V and X).  If we use an MPHF $h$ with $h (5) = 1$, $h (3) = 2$ and $h (2) = 0$, then we can sort the non-terminals into the order V, X; Z; W, Y, with semicolons showing the divisions between the groups.

Assuming the alphabet is $\{ \mbox{\$}, \mbox{A}, \mbox{C}, \mbox{G}, \mbox{T} \}$, we replace the non-terminals by the triples $(1, 1, 4), (1, 4, 1); (3, 0, 1); (1, 3, 0), (1, 2, 0)$, with the semicolons again showing the divisions between the groups. For example non-terminal V is represented by $(1,1,4)$ since its left child, the terminal A, has expansion size 1, and its offset among the terminals is 1, while the second child, the terminal T, has offset 4.
Finally, we encode the rule involving the initial symbol S as the bitvector $0000100110011000010011001$, which is the concatenation of the unary representations of the expansion sizes of the symbols on the rule's right-hand side, and the sequence $0, 0, 1, 1, 0, 0, 1, 1, 0$ giving the offset of each symbol in its group. 

To extract the 17th character of the text, we start by performing a rank query and two select queries on the bitvector for S, which together tell us that the 17th character is the 4th character in the expansion of the 6th symbol on the right-hand side of the rule for S, and that symbol expands into 5 characters.  Checking the sequence for S, we see that the 6th symbol on the right-hand side of the rule for S has rank 0 among all the non-terminals that expand to 5 characters (note there is only one such non-terminal, Z).

We compute $h (5) = 1$ and check the triple with rank 0 in the group with rank 1 --- i.e., $(3, 0, 1)$ --- which tells us that Z's left child expands into 3 characters, so its right child X expands into 2 characters and the 4th character in the expansion of Z is the 1st character in the expansion of X, and that X has rank 1 among the non-terminals that expand into 2 characters.  Note that we never actually learn or use the identifiers Z or X in the actual data structure: we use them here just to ease the presentation.  We compute $h (2) = 0$ and check the triple with rank 1 in the group with rank 0 --- i.e., $(1, 4, 1)$ --- which tells us that X's left child expands into 1 character, so it is a terminal, and it has rank 4 in the alphabet, meaning it is a T.

Admittedly, for this small example we do not save space compared to the na\"ive encoding, but our experiments show that it pays to carefully integrate our encodings of the symbols in the parse and its shape.

\section{Experiments}
\label{sec:experiments}

We compared our encoding with the na\"ive encoding and the state-of-the-art encoding by Maruyama et al.~\cite{MaruyamaTSS13}; we refer to these as {\sf OURS}, {\sf NAIVE} and {\sf MTSS}, respectively.  For the na\"ive encoding of an SLP for a string of length $n$ with $r$ rules, we store the following information in plain arrays:
\begin{enumerate}
    \item the right-hand sides of rules in $2 r \lg (r + \sigma)$ bits,
    \item the expansion length for every non-terminal in $r \lg n$ bits.
\end{enumerate}
To support random access to the triples in our encoding and to store the bitvector for the start rule, we used SD bitvectors from the SDSL 2.0 library\footnote{\url{https://github.com/simongog/sdsl-lite}}.  Our experiments ran on a Xeon E5-1650V3 (6core/12thread 3.5GHz) machine with 32 GB memory.

In this section we describe only our main experimental results; additional results can be found in Appendix~\ref{app:additional}.  For our main experiments, we used the same 59 GB file containing 1000 copies of chromosome 19 that we used in our previous work~\cite{GIMNST19}, downloaded from the 1000 Genomes Project~\cite{1000Genomes}; the effective alphabet size was 5.  When we compress the dictionary and parse with Navarro's implementation of RePair combined with CTPH, as described in Section~\ref{sec:background}, the resulting 59 MB SLP contains almost 13 million rules with almost 120\,000 distinct expansion lengths and almost 4.5 million symbols on the right-hand side of the start rule; the height of the parse tree is 43.

Table~\ref{tab:results} shows our main experimental results: for each of the given substring lengths and each of the encodings, we extracted that many consecutive characters from 10000 pseudo-randomly chosen positions in the compressed file and averaged the extraction times.  The na\"ive encoding is obviously the largest but also the fastest: it takes 217 MB,  access to a single character taking 1.8 microseconds, and access to ten consecutive characters taking 2.2 microseconds.  Maruyama et al.'s encoding takes 86 MB --- much closer to the size of the unaugmented SLP --- but access to one character takes 26 microseconds and access to ten takes 30 microseconds.  We encode the augmented grammar in 81 MB --- even less than Maruyama et al. --- with access to one character taking 6.9 microseconds and access to ten taking 9.3 microseconds.  Although our encoding is still significantly slower than the na\"ive encoding, it is only a little more than a third of the size.  The size difference is particularly pronounced if we compare how much larger the na\"ive encoding and ours are than the unaugmented SLP: $217 / 59 \approx 3.7$ versus $81 / 59 \approx 1.4$.  Building our encoding is also reasonably fast, taking only 18 seconds with the source code we have made publicly available at {\tt https://github.com/itomomoti/ShapedSlp}\,.

\begin{table}[t]
    \centering
    \begin{tabular}{rrrr}
    \multicolumn{1}{c}{substring} & \multicolumn{1}{c}{\sf NAIVE} & \multicolumn{1}{c}{\sf MTSS} & \multicolumn{1}{c}{\sf OURS}\\
    \multicolumn{1}{c}{length} &  \multicolumn{1}{c}{(217 MB)} &  \multicolumn{1}{c}{(86 MB)} & \multicolumn{1}{c}{(81 MB)}\\
    \hline
         1 &  1.8 &  25.9 &   6.9\\
        10 &  2.2 &  29.6 &   9.3\\
       100 &  5.2 &  63.5 &  31.7\\
      1000 & 31.6 & 394.6 & 249.6
    \end{tabular}
    \medskip
    \caption{Extraction times in microseconds with the three encodings and various substring length.}
    \label{tab:results}
\end{table}

\begin{figure}[p!]
\centering
\includegraphics[scale=0.4]{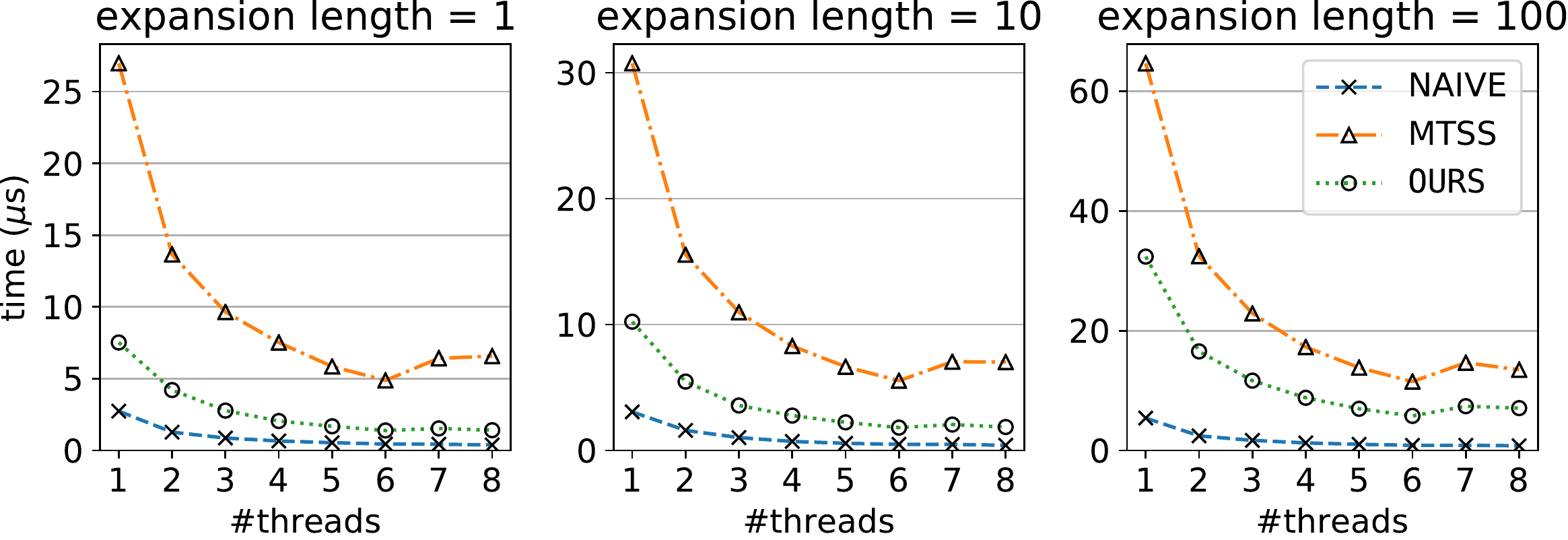}
\caption{Average time to answer an expansion query using multiple threads.}
\label{fig:chr19q}
\end{figure}

For some applications, we are interested in processing many queries at once, which offers us the opportunity to exploit parallelism.  Figure~\ref{fig:chr19q} shows the average speedup using up to 8 threads.  Since the scale makes it difficult to discern the height of the rightmost points, we note that {\sf NAIVE}, {\sf MTSS} and {\sf OURS} with 8 threads use 0.38, 6.56 and 1.41 microseconds for length 1; 0.41, 7.01 and 1.86 for length 10; and 0.78, 13.47 and 7.07 for length 100.

\subsubsection*{Acknowledgements:}

TG was partly funded by NSERC RGPIN-2020-07185, Canada, and Basal Funds FB0001, Chile.  TI, HS and YT were partly funded by JSPS KAKENHI grants 19K20213, 17H01791 and 18K18111, respectively.  GM was partly funded by MIUR-PRIN grant 2017WR7SHH.  GN was partly funded by Basal Funds FB0001 and Fondecyt grant 1-200038, Chile.  LSB was partly funded by DFG project LO 748/10-2 (QUANT-KOMP) and received travel funds from the EU's Horizon 2020 MSC RISE program (grant 690941).

\appendix

\section{Theoretical bounds}
\label{app:theory}

Charikar et al.~\cite{CharikarLLPPSS05} and Rytter~\cite{Rytter03,Rytter04} independently showed how, given a text $T$ of length $n$ over an alphabet of size $\sigma$ whose smallest SLP has $g^*$ rules, in $O (n \log \sigma)$ time we can build an SLP for $T$ with $O (g^* \log (n / g^*))$ rules and height $O (\log n)$.  We can augment the non-terminals of this SLP with the sizes of their expansions to obtain an $O (g^* \log (n / g^*))$-space data structure supporting access to any $\ell$ consecutive characters of $T$ in $O (\log n + \ell)$ time.  Bille et al.~\cite{BilleLRSSW15} showed how we can take any SLP for $T$ with $g$ rules, regardless of height, and build a data structure of size $O(g)$ (measured in words of bit length $\log n$) that also supports access to any $\ell$ consecutive characters in $O (\log n + \ell)$ time, while Verbin and Yu~\cite{VerbinY13} proved we generally cannot support $O (\log^{1 - \epsilon} n)$-time random access to $T$ with a $\mathrm{poly} (g)$-space data structure.  Belazzougui et al.~\cite{BelazzouguiCPT15} showed how we can support $O (\log n / \log \log n)$-time random access to $T$ with an $O (g \log^\epsilon n)$-space grammar.  Prezza~\cite{Prezza19} sidestepped Verbin and Yu's lower bound to obtain constant-time random access to $T$ with an $O (g n^\epsilon)$-space grammar (after Belazzougui et al.~\cite{belazzougui2015queries} achieved that tradeoff with block trees).  Recently, Ganardi, Je\.{z} and Lohrey~\cite{GanardiJL19} showed how we can turn any SLP for $T$ with $g$ rules into an SLP for $T$ with $O (g)$ rules and height $O (\log n)$, thus simplifying many previous proofs.

Regarding SLPs produced with RePair, Charikar et al.~\cite{CharikarLLPPSS05} showed they can be an $\Omega (\log^{1 / 2} n)$ factor larger than the smallest possible SLPs, and Hucke, Je\.{z}  and Lohrey~\cite{HuckeJL17,BHHIJLR19} improved that lower bound to $\Omega (\log n / \log \log n)$.  Charikar et al. showed they are always within an $O ((n / \log n)^{2 / 3})$-factor of the smallest SLPs and this is still the best upper bound known, although Hucke~\cite{Huc19} showed they are within a $\log_2 3$-factor for unary strings.

\section{Additional experimental results}
\label{app:additional}

We are mainly interested in compressing human DNA but we performed experiments with other datasets to check our approach's robustness: 11264 Salmonella genomes (salx11264) from the GenomeTrakr project~\cite{STBASBM17}, and two repetitive files from the Pizza \& Chili corpus\footnote{\url{http://pizzachili.dcc.uchile.cl/}} (einstein.en.txt and kernel).

\begin{figure}[p!]
\centering
\includegraphics[scale=0.4]{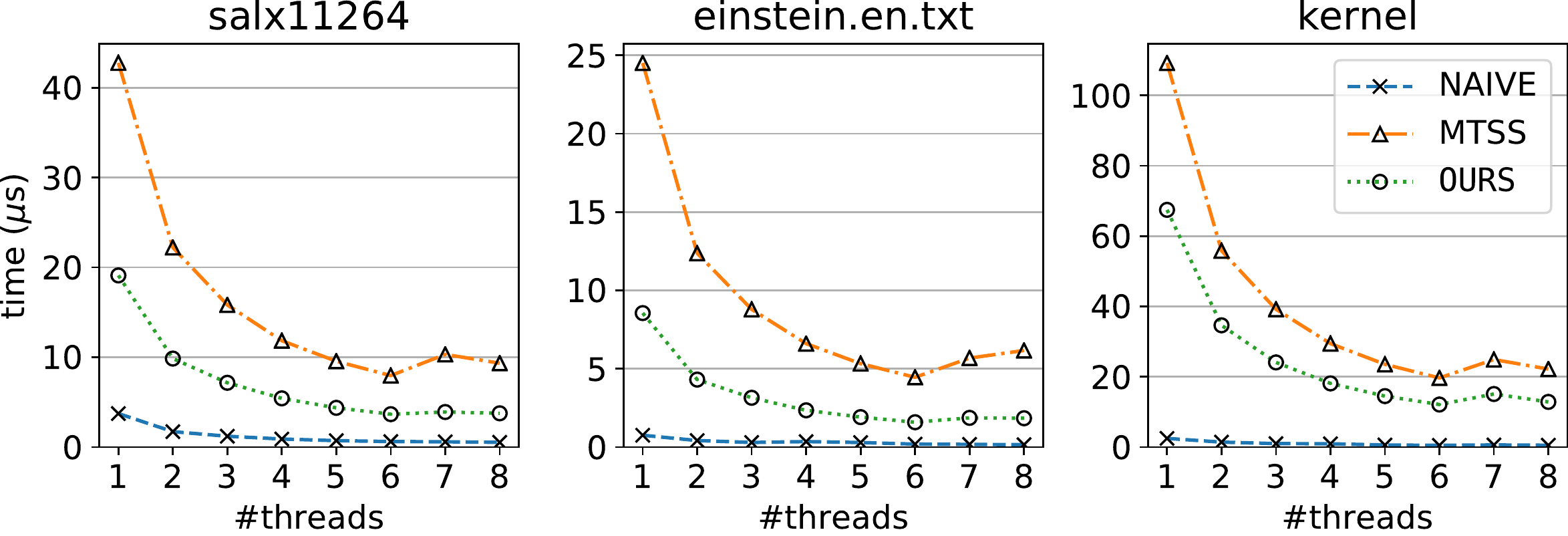}
\caption{Average time to answer an expansion query with expansion length 10 using multiple threads.}
\label{fig:various}
\end{figure}

As can be seen from the tables below and comparing Figure~\ref{fig:chr19q} to Figure~\ref{fig:various}, our results are not as good for the other datasets as for chr19x1000 but the our general conclusions are supported: {\sf MTSS} and {\sf OURS} are about the same size and several times smaller than {\sf NAIVE}; {\sf NAIVE} is by far the fastest to build, with {\sf MTSS} slower by almost an order of magnitude and {\sf OURS} slower even than that by a factor of 4 to 7; {\sf NAIVE} is also the fastest to answer queries, followed by {\sf OURS} and then {\sf MTSS}.  Since the scale again makes it difficult to discern the height of the rightmost points, we note that {\sf NAIVE}, {\sf MTSS} and {\sf OURS} with 8 threads use 0.53, 9.34 and 3.76 microseconds for salx11264; 0.15, 6.16 and 1.84 for einstein.en.txt; and 0.53, 22.18 and 12.84 for kernel.

\begin{table}[p!]
    \centering
    \begin{tabular}{lrrrrrr}
    dataset & $\sigma$ & $n$ & $s$ & $r$ & $d$ & $h$\\
    \hline
    chr19x1000 & 5 & 59125115010 & 4495360 & 12898128 & 118889 & 43\\
    salx11264 & 4 & 57033515255 & 32579379 & 199121788 & 332808 & 18658\\
    einstein.en.txt & 139 & 467626544 & 62473 & 100611 & 17343 & 1353\\
    kernel & 160 & 257961616 & 69427 & 1057914 & 48453 & 5820\\
    \end{tabular}
    \smallskip
    \caption{Statistics of our datasets: name, alphabet size, length (in bytes), number of symbols on the right-hand side of the start rule, number of rules, number of distinct expansion lengths, and height of the grammar.}
    \label{tab:stats}
\end{table}

\begin{table}[p!]
  \centering
  \begin{tabular}{lrrrrrr}
    &  \multicolumn{3}{c}{size (bytes)} & \multicolumn{3}{c}{construction time (ms)} \\
    dataset & \sf NAIVE & \sf MTSS & \sf OURS & \sf NAIVE & \sf MTSS & \sf OURS\\
    \hline
    chr19x1000 & 217418909 & 86362255 & 80629662 & 524 & 4576 & 17649\\
    & (0.37\%) & (0.15\%) & (0.14\%) &&&\\
    salx11264 & 2896264885 & 799395665 & 956575138 & 5457 & 53147 & 370175\\
    & (5.1\%) & (1.4\%) & (1.7\%) &&&\\
    einstein.en.txt & 1896040 & 674979 & 631698 & 3 & 22 & 92\\
    & (0.41\%) & (0.14\%) & (0.14\%) &&&\\
    kernel & 12964629 & 4473636 & 5044020 & 30 & 158 & 866\\
    & (5.0\%) & (1.7\%) &	(2.0\%) &&&
    \end{tabular}
    \smallskip
    \caption{Sizes of the encodings and construction times.}
    \label{tab:size}
\end{table}

\end{document}